\documentclass[final,5p,times]{elsarticle}

 \usepackage{graphicx}
 \usepackage{subfig}

\usepackage{amssymb}

 \usepackage{lineno}

\journal{Nuclear Physics B}

\begin{document}

\begin{frontmatter}



\title{A new CMS pixel detector for the LHC luminosity upgrade}


\author{Carlotta Favaro \textit{on behalf of the CMS Collaboration}}

\address{Universit\"at Z\"urich, Physik-Institut, Winterthurerstr. 190, 8057 Z\"urich, Switzerland}

\begin{abstract}
The CMS inner pixel detector system is planned to be replaced during the first phase of the LHC luminosity upgrade. The plans foresee an ultra low mass system with four barrel layers and three disks on either end. With the expected increase in particle rates, the electronic readout chain will be changed for fast digital signals. An overview of the envisaged design options for the upgraded CMS pixel detector is given, as well as estimates of the tracking and vertexing performance. 
\end{abstract}


\end{frontmatter}


\section{Motivations for the Phase 1 upgrade of the pixel detector}
\label{motivations}
The silicon pixel detector \cite{pixeldet} is the innermost part of CMS. It has the key role to provide the precise spatial measurements used as seeds for the reconstruction of charged particle trajectories in proximity of the primary interaction point. Its performance is thus crucial for the identification of primary and secondary vertices, and for the measurement of long-lived particles such as $b$ quarks and $\tau$ leptons. The present detector was designed for a maximum peak luminosity of 10$^{34}$ cm$^{-2}$s$^{-1}$, the design value of the Large Hadron Collider, which will be exceeded in the so called Phase 1. Higher luminosities are not sustainable, mostly due to readout inefficiencies. \\
To fully profit from the large datasets that will be collected by the CMS experiment, a new optimized silicon pixel detector will be installed during the long LHC shutdown in 2016.

The new detector will be required to retain a good hit detection efficiency and prevent data losses in the large occupancy environment, to assure good track seeding and pattern recognition performance, and to provide high resolution on track parameters.\\
All modifications will be constrained by the existing cables and off-detector services, since space limitations prevent the installation of additional components. Moreover, the number of detector module types will have to be reduced, in order to limit the time and costs for production and testing.

In addition, despite the use of radiation resistant technologies, the detector is not sufficiently radiation hard to survive until the end of the Phase 1, when a total integrated luminosity of 350 fb$^{-1}$ will be collected. The innermost barrel layer will have sustained a particle fluence of about 10$^{16}$ n$_{eq}$cm$^{-2}$, producing an irreversible degradation of its performance. An intermediate replacement of this layer will therefore be needed. 

\section{Detector design and material budget}
\label{design} 
\begin{figure}[htbp]
\begin{center}
\includegraphics[height=5.2cm]{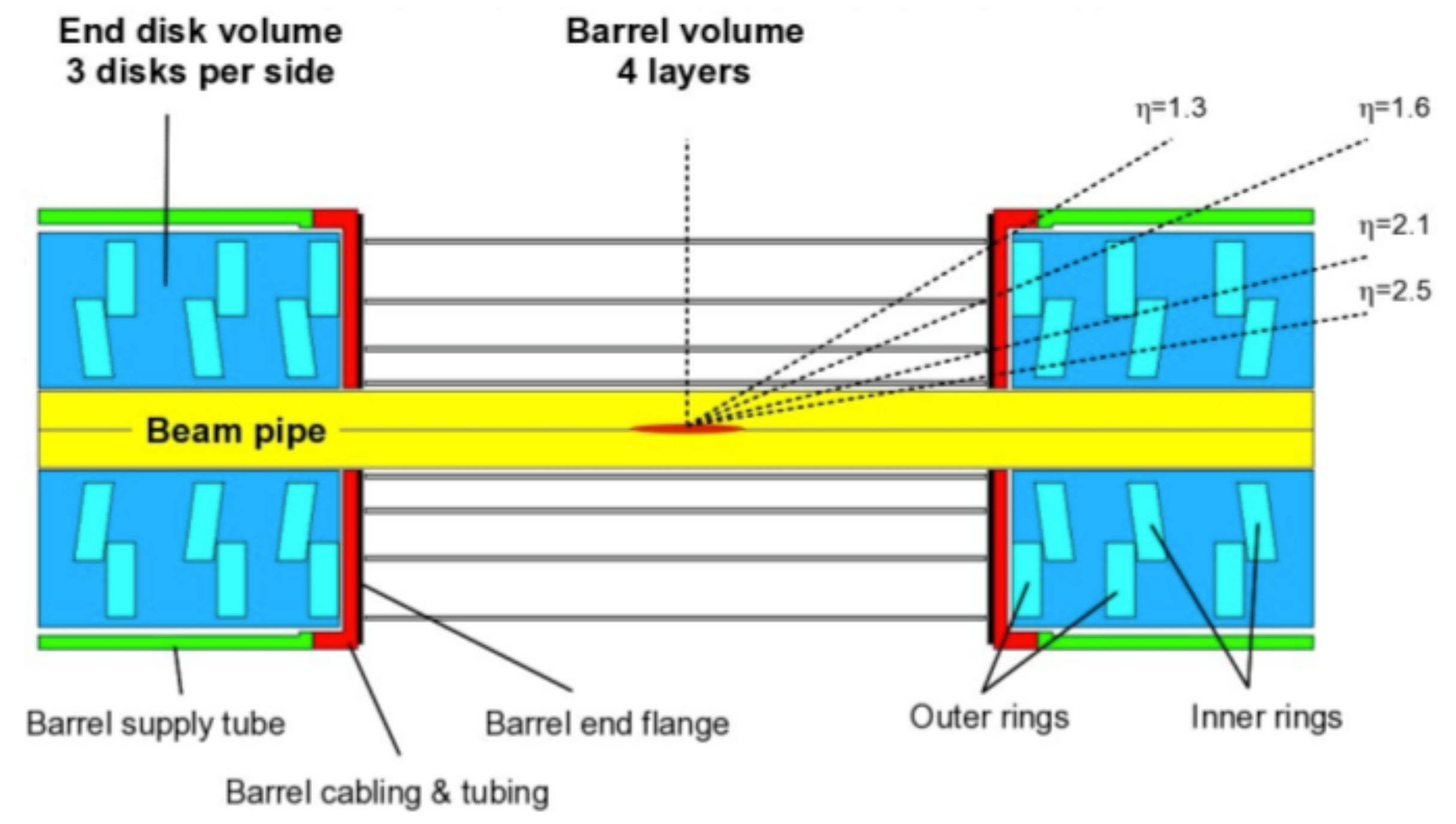}
\caption{\small{Longitudinal cross section of the upgraded pixel detector, showing the location of barrel layers and endcap disks.}}
\label{det}
\end{center}
\end{figure}

The present pixel detector does not provide a hermetic three-hit coverage. The resulting seeding inefficiencies limit the performance of the High Level Trigger, and slow the offline reconstruction, based on a sophisticated iterative algorithm \cite{tracking}.\\
 A new geometrical layout, shown in Fig.\ref{det}, will thus be implemented.
The proposed barrel design includes four cylindrical layers, placed at radii of 3.9, 6.8, 10.9 and 16.0 cm. The innermost layer is moved closer to the interaction point, while the forth layer is added in the gap between the present third pixel and the first strip layers. This will result in a factor two increase of the radial acceptance and a reduction of the extrapolation distance between pixel and strip detectors, with a benefit to the pattern recognition. \\
Three endcap disks will be installed at each side of the barrel, at 29.1, 39.6 and 51.6 cm from the interaction point. The new layout will provide an almost hermetic four-hit coverage up to a pseudorapidity of 2.5. 

One module type will be used in both barrel and endcap regions. Each module includes a silicon pixel sensor \cite{sensor} bump-bonded to 16 readout chips. In the barrel, the modules will be mounted on carbon fiber ladders glued onto stainless steel cooling tubes. In the endcaps the support structure will be made of blades arranged radially into half-disks, with a similar turbine-like geometry of the present detector. Each half-disk will be composed of two concentrical rings, to remove and replace independently the innermost part after radiation damage. 

\begin{figure}
  \centering
   \subfloat{
     \label{materialA}           
       \includegraphics[height=4.4cm]{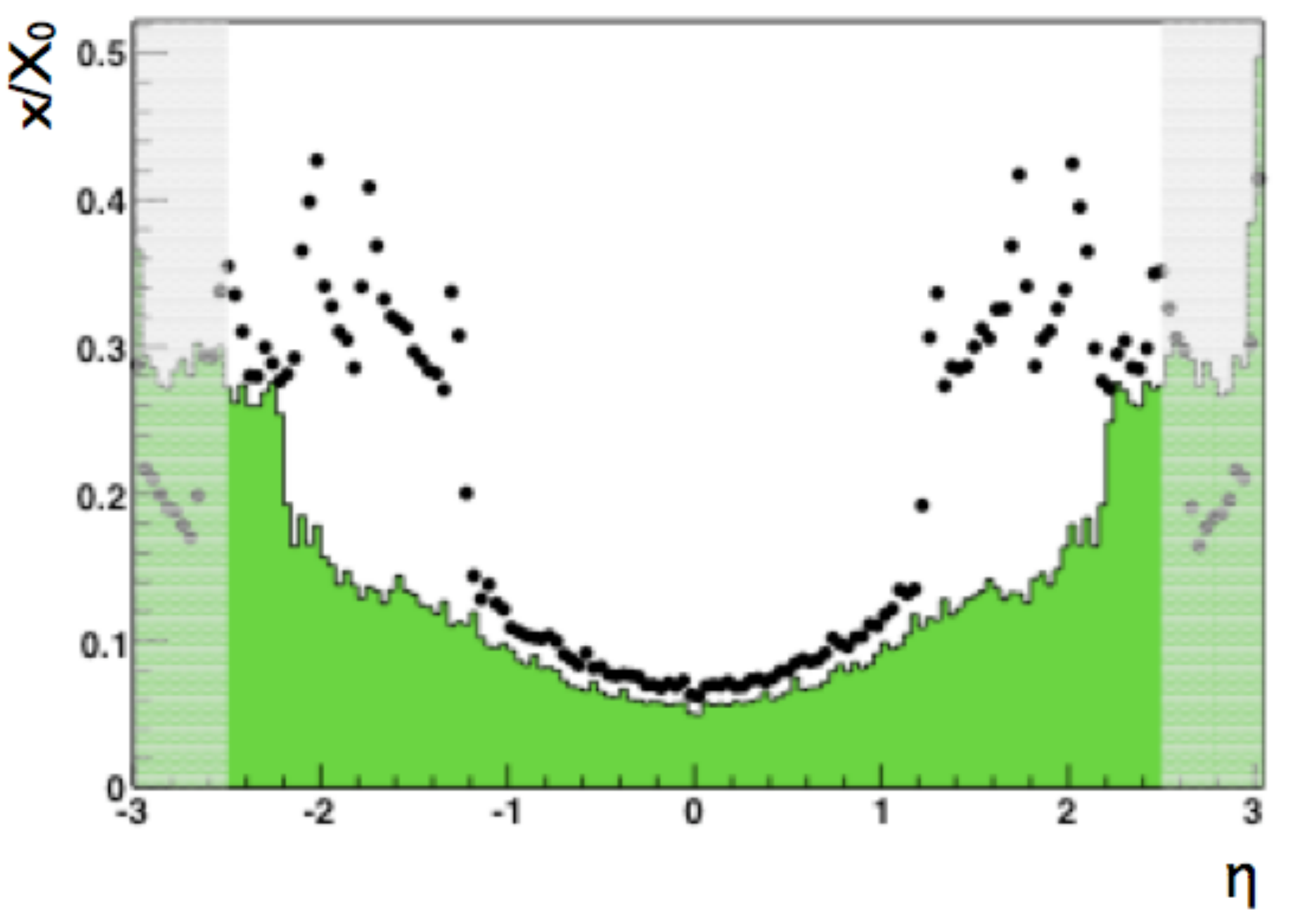}}
   \hspace{0.1cm}
   \subfloat{
     \label{materialB}           
        \includegraphics[height=4.4cm]{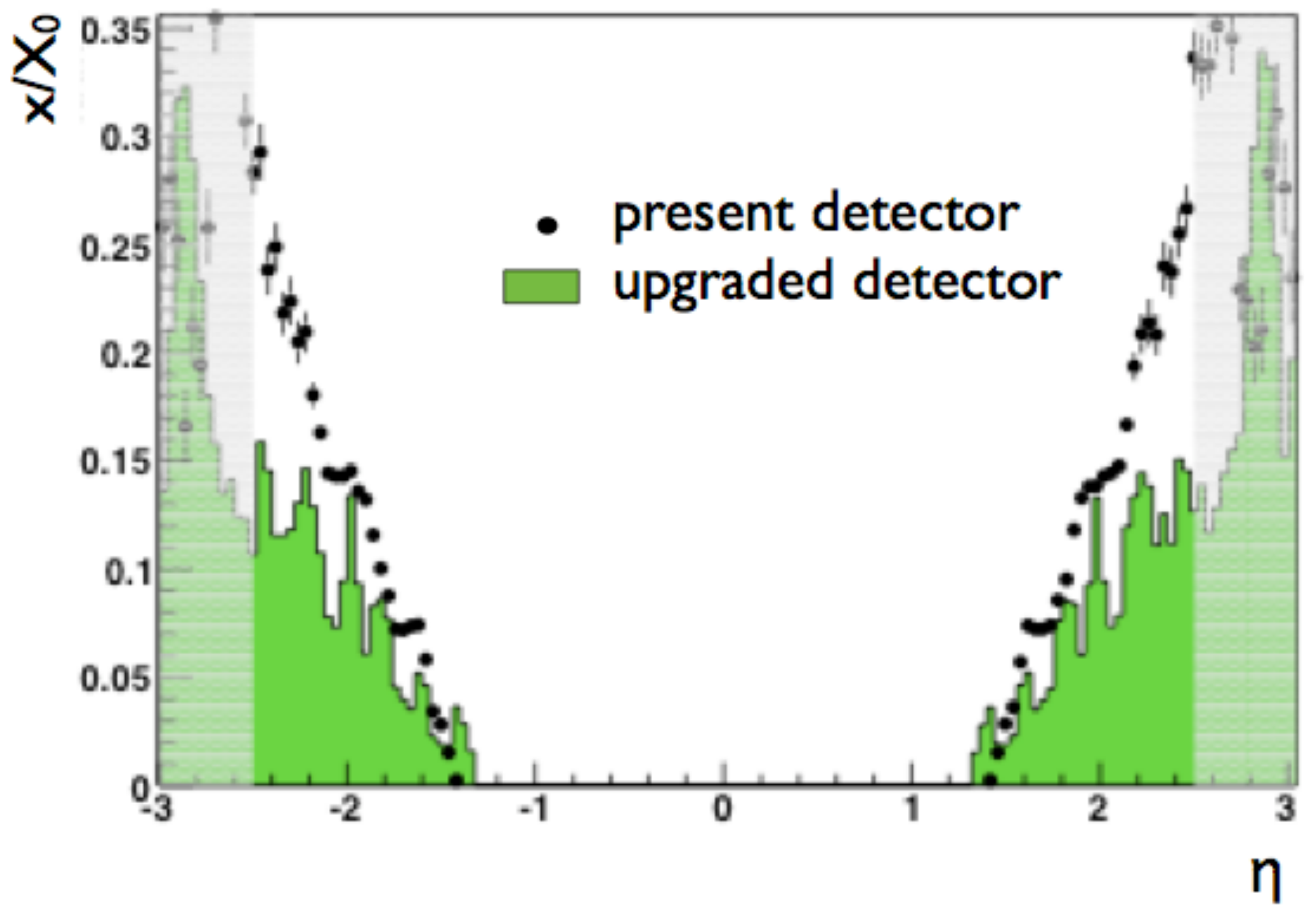}}
   \caption{\small{Material budget of the barrel (top) and the endcap (bottom) detector in term of radiation lengths for the present (black dots) and the upgraded (green histogram) systems, as a function of track pseudorapidity. The grey bands show the pseudorapidity region outside the tracking acceptance. }}
   \label{material}                  
\end{figure}

A limitation of the present pixel detector is the significant amount of material within the tracking acceptance, which degrades the performance of track reconstruction. The biggest contributions are the silicon sensors, the mechanical support, the cooling system, and electronics. In addition, the barrel endflange hosting cooling manifolds and electronic boards is a considerable amount of material located in front of the first forward disk. One of the objectives of the Phase 1 upgrade is a drastic reduction of the material budget. The present C$_{6}$F$_{14}$ will be replaced by a two-phase CO$_{2}$ system, which has suitable thermodynamic properties for flowing in micro-channels, low mass and sufficient radiation hardness. In both barrel and endcap the modules will be installed on ultra-lightweight support structures. Most part of the barrel services currently on the endflange will be moved to the barrel supply tube, outside the tracking acceptance. \\
These modifications will result in at least a factor of two reduction of the material budget, as shown in Fig.\ref{material}.

Finally, a new powering system will be needed to power the increased number of components with the present cables and services. DC-DC converters will be used for this purpose. 

\section{Readout}
\label{readout}
The present pixel readout chip (PSI46v2) \cite{roc} was designed to provide high hit detection efficiency at the design LHC peak luminosity of 10$^{34}$ cm$^{-2}$s$^{-1}$. Assuming a Level 1 Trigger accepted rate of 100 kHz, at the instantaneous luminosity of 2$\times$10$^{34}$ cm$^{-2}$s$^{-1}$ the dynamic inefficiency of the innermost barrel layer is estimated to grow from 4\% to 16\%, with unacceptable deterioration of hit reconstruction and track seeding efficiencies.

The main sources of readout inefficiency must be addressed to prevent data losses. In order to keep a high single-hit efficiency, the size of the data buffers at the double column periphery will be extended from the present 32 to 80 units, compatibly with space limitations. An additional buffer stage will be introduced, to store the Level 1 Trigger accepted hit information while waiting for the readout token. \\
Moreover, the implementation of a faster readout will be needed to read the increased number of channels through the existing optical fibers. The plan is to switch from the current analogue signal to digital, with an on-chip ADC. In addition high speed (320 Mbps) links will be used to increase the bandwidth.

With these modifications the dynamic inefficiency is estimated to be about 5.7\% for a trigger rate of 100 kHz and a peak luminosity of 2$\times$10$^{34}$ cm$^{-2}$s$^{-1}$.  
\begin{figure}
   \centering
     \subfloat{
        \label{IP_transv}           
        \includegraphics[height=4.2cm]{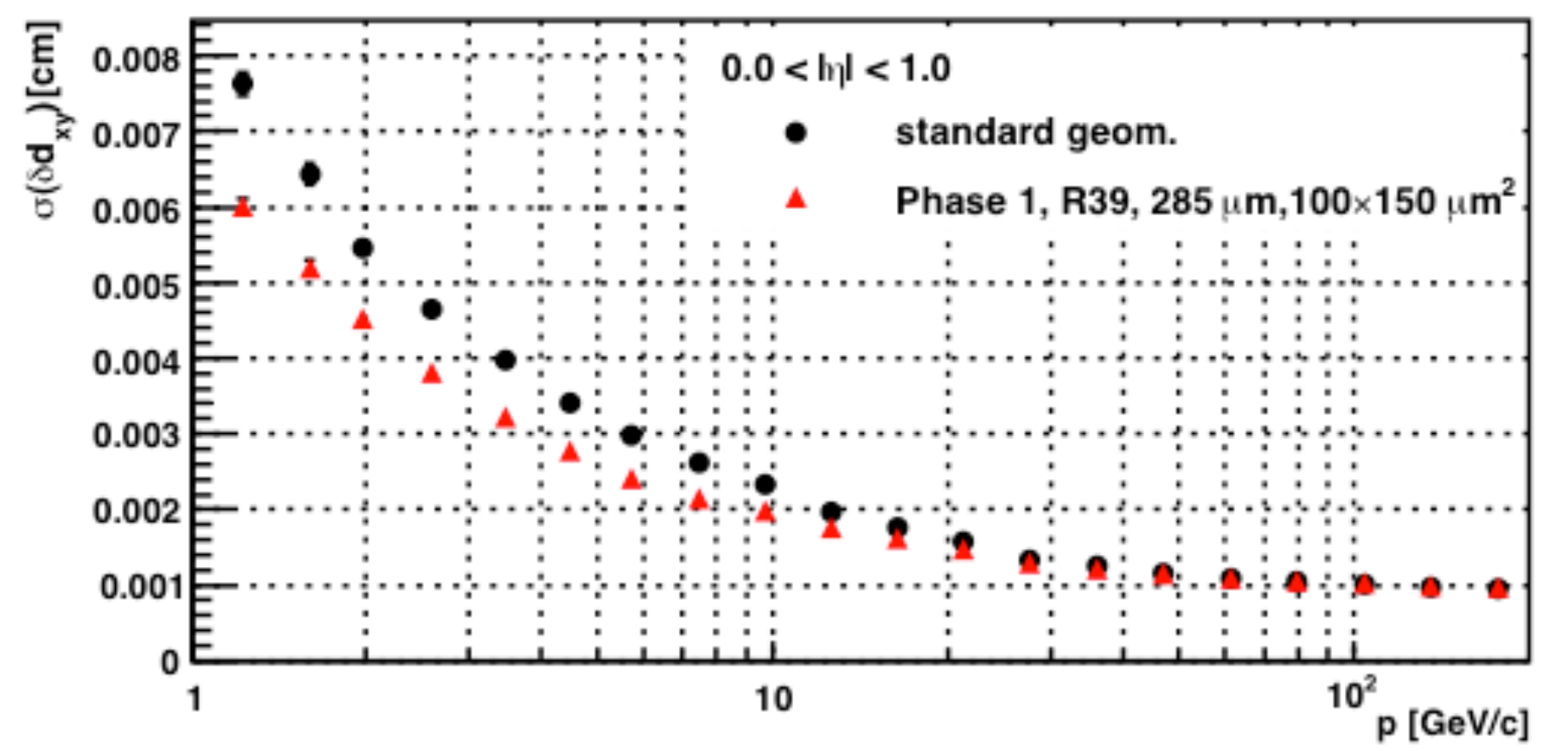}}
   \hspace{0.1cm}
   \subfloat{
     \label{IP_long}           
        \includegraphics[height=4.2cm]{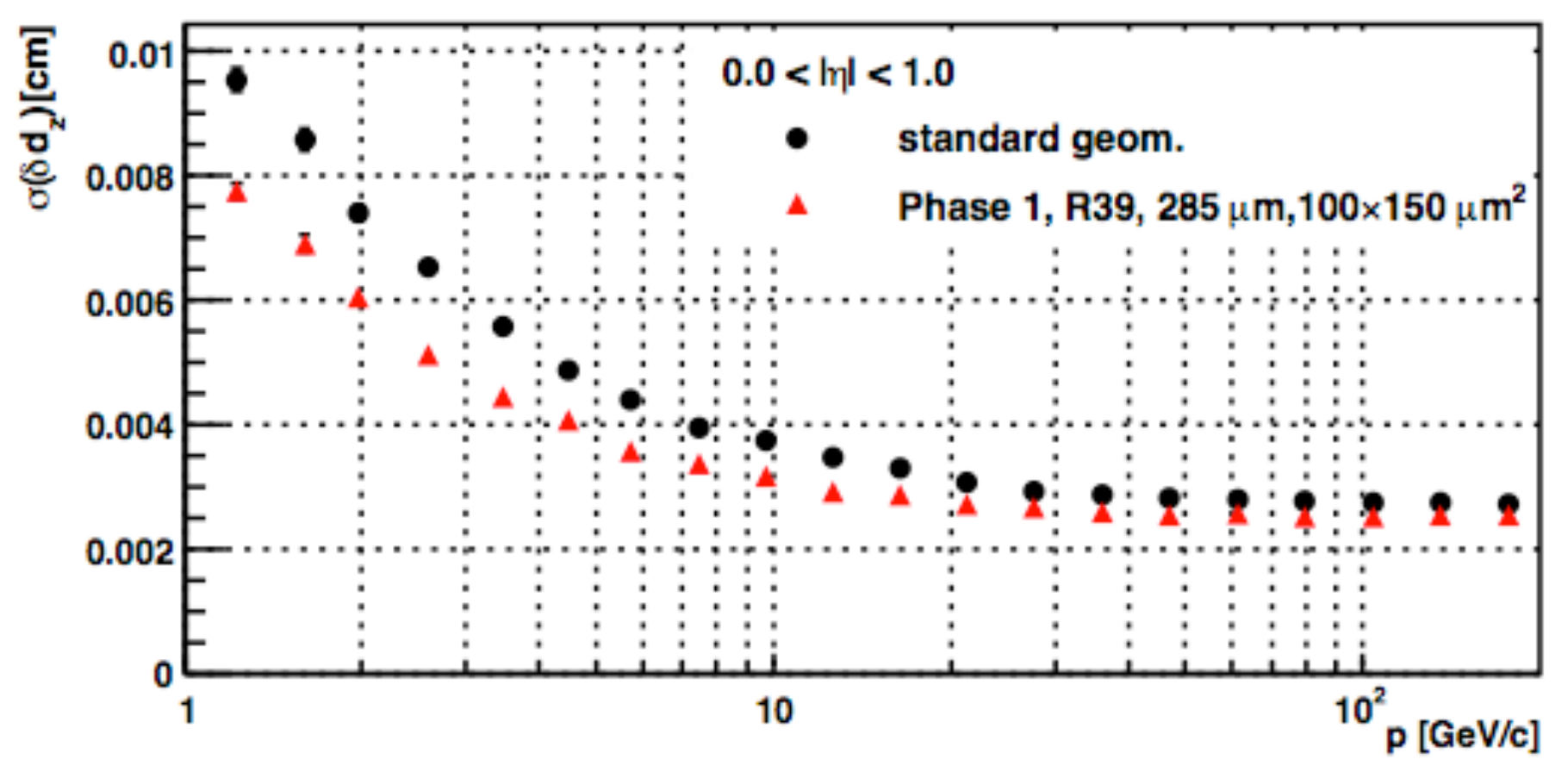}}
   \caption{\small{ Transverse (top) and longitudinal (bottom) impact parameter resolution for present (black) and upgraded (red) pixel detectors as functions track momentum, for muon events. 
   The baseline upgraded configuration with the innermost layer at a radius of 39 mm, a sensor thickness of 285 $\mu$m and a pixel cell size of 100$\times$150 $\mu$m$^{2}$ was considered. Only the barrel is shown. The improvement in the forward region is about 40\%.}}
   \label{IP}                  
\end{figure}

\section{Performance improvement}
\label{performance}
 The enhanced features of the new pixel detector will produce a significant improvement of the performance, in terms of track parameters resolution, tracking efficiency and fake rate, vertex reconstruction and $b$-tagging.

The material reduction, the increase of the radial acceptance, and the four-hit hermetic coverage will provide a better resolution of the track parameters. The improvement will affect both fully reconstructed and pixel-only tracks, used by the High Level Trigger. \\
\begin{figure}
   \centering
     \subfloat{
        \label{PV_transv}           
        \includegraphics[height=4.5cm]{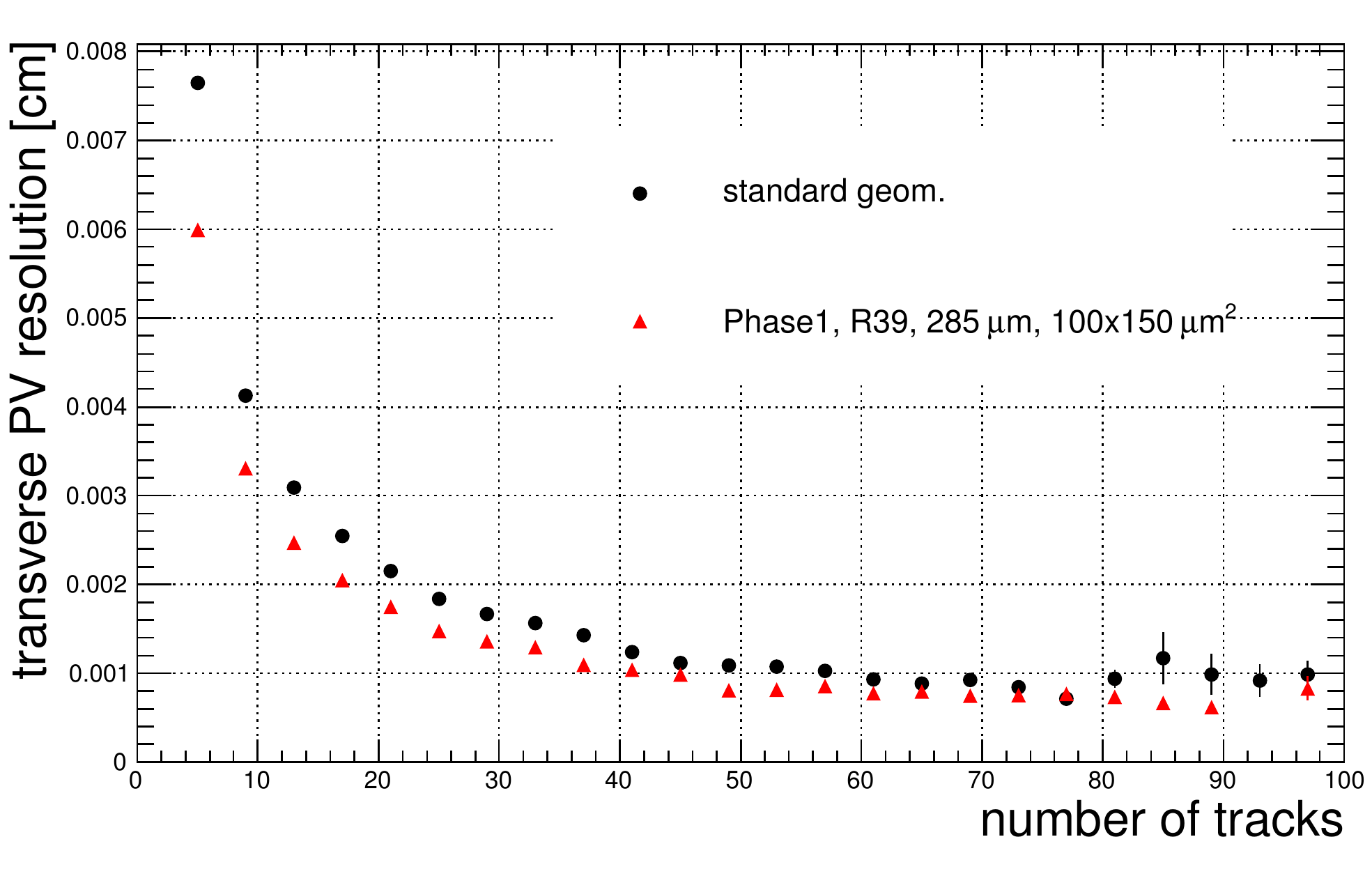}}
   \hspace{0.1cm}
   \subfloat{
     \label{PV_long}           
        \includegraphics[height=4.5cm]{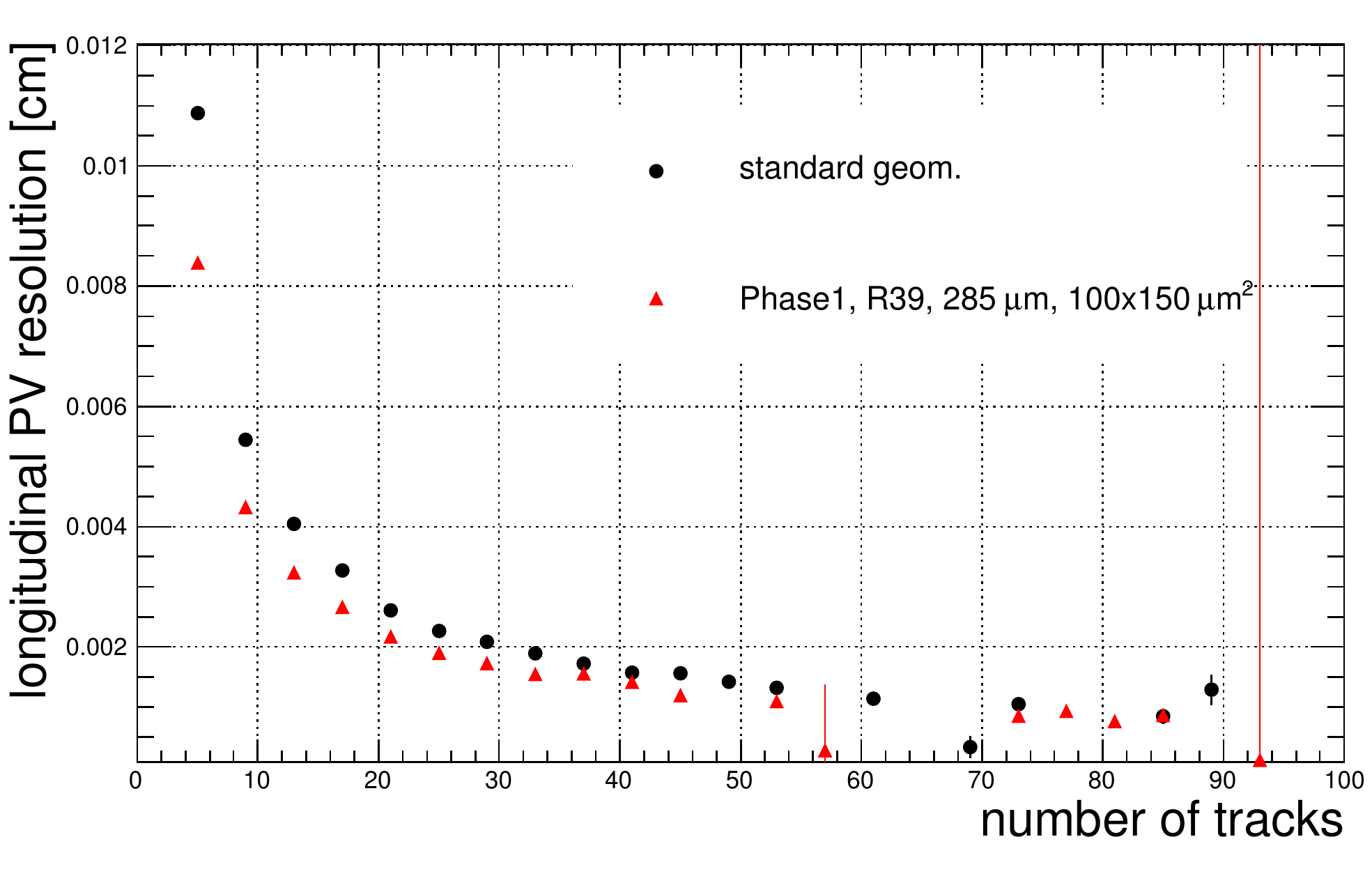}}
   \caption{\small{ Transverse (top) and longitudinal (bottom) primary vertex resolution for present (black) and upgraded (red) pixel detectors as functions of number of tracks, for a sample of top events at the instantaneous luminosity of 10$^{34}$ cm$^{-2}$s$^{-1}$, from full simulation.  The baseline upgraded configuration with the innermost layer at a radius of 39 mm, a sensor thickness of 285 $\mu$m and a pixel cell size of 100$\times$150 $\mu$m$^{2}$ was considered. The improvement in the both cases is about 20\%.}}
   \label{PV}                  
\end{figure}
Fig.\ref{IP} shows the transverse and longitudinal impact parameter resolution of fully reconstructed tracks, for the present and upgraded detectors. Only the result for the barrel is presented. The improvement will be about 25\%, and will reach 40\% in the forward region in correspondence of the location of the endflange. The effect is more pronounced in the low momentum region, where the multiple scattering is dominant and the sensitivity to the material reduction is thus bigger. A factor of four enhancement is also expected for pixel-only tracks, thanks to the extended radial coverage.

The better track parameter resolution will enhance the vertex reconstruction \cite{tracking,vertex} performance. In Fig.\ref{PV} the primary vertex resolution is shown as a function of the number of tracks associated to the vertex, at the peak luminosity of 10$^{34}$ cm$^{-2}$s$^{-1}$, for the present and the upgraded detectors. The improvement is about 20\% in both transverse and longitudinal planes. This will be crucial to disentangle multiple interactions, 25 on average at  10$^{34}$ cm$^{-2}$s$^{-1}$, within a bunch crossing. Since a similar effect is expected for displaced secondary vertices, a 20\% improvement of lifetime measurements is also foreseen. 

\begin{figure}[htbp]
\begin{center}
\includegraphics[height=5.6cm]{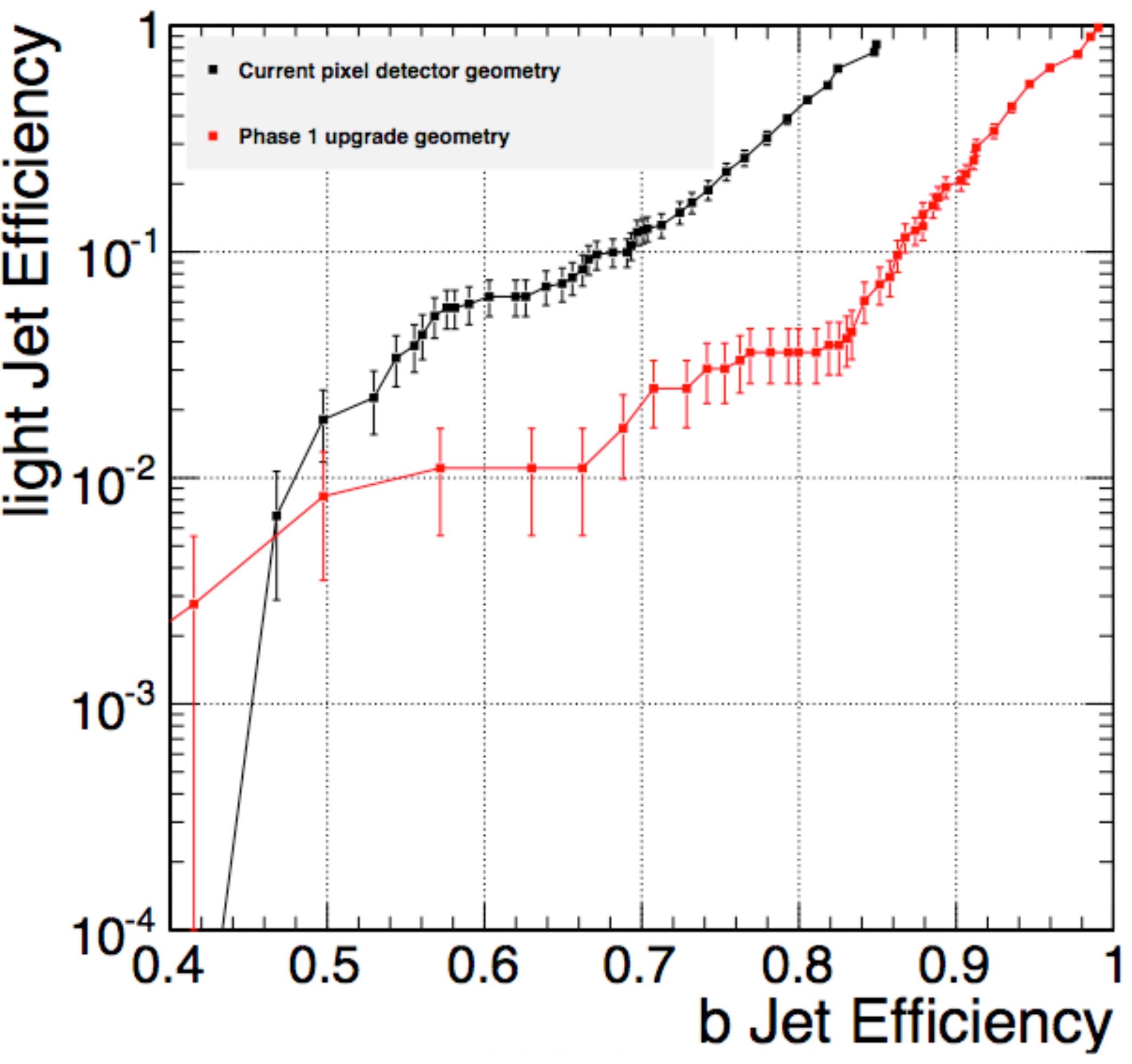}
\caption{\small{$b$-tagging efficiency and contamination from light quarks for the Vertex algorithm \cite{btag}, for the present (black) and the upgraded (red) setups. A sample of  top events at an instantaneous luminosity of 10$^{34}$ cm$^{-2}$s$^{-1}$, obtained with full simulation, was used.}}
\label{B}
\end{center}
\end{figure}

The enhancement of both tracking and vertexing performance will improve the identification of $b$-jets. The Vertex tagger \cite{btag} was tested on a sample of $t \overline{t}$ events, at a peak luminosity of 10$^{34}$ cm$^{-2}$ s$^{-1}$. The result is shown in Fig.\ref{B}. The new system will provide a factor 6 reduction of the contamination from light quark jets with a 20\% increase of efficiency.

The four-hit coverage will offer the opportunity to implement a tracking algorithm with quadruplet seeding. This will produce an increase of the efficiency with a reduction of the fake rate, which will be crucial in the dense hit environment.

The pattern recognition will be faster and less affected by the combinatorics, thanks to the bigger radial acceptance and to the smaller gap between the outermost pixel and the innermost strip layer.

Finally, the material decrease will reduce the photon conversion, leading to improved electron reconstruction, and will reduce the rate of secondary tracks from nuclear interactions.

\section{Further development for late Phase 1}
\label{latephase1}
The innermost part of the pixel detector is expected to be heavily degraded before the end of the Phase 1 \cite{radhardnessBPIX,radhardnessFPIX}. 

All components of the present system were designed to operate up to a total particle fluence of 6$\times$10$^{14}$ n$_{eq}$cm$^{-2}$. Two years of operation at the late Phase 1 luminosity are equivalent for the innermost layer of the barrel, and for the endcap at similar radii, to a fluence of 10$^{16}$ n$_{eq}$ cm$^{-2}$. At this dose the performance of the silicon sensor will be heavily degraded. Charge collection was measured in test beams as a function of the bias voltage, for various radiation fluences \cite{radhardnessBPIX}. 
\begin{figure}[htbp]
\begin{center}
\includegraphics[height=5.0cm]{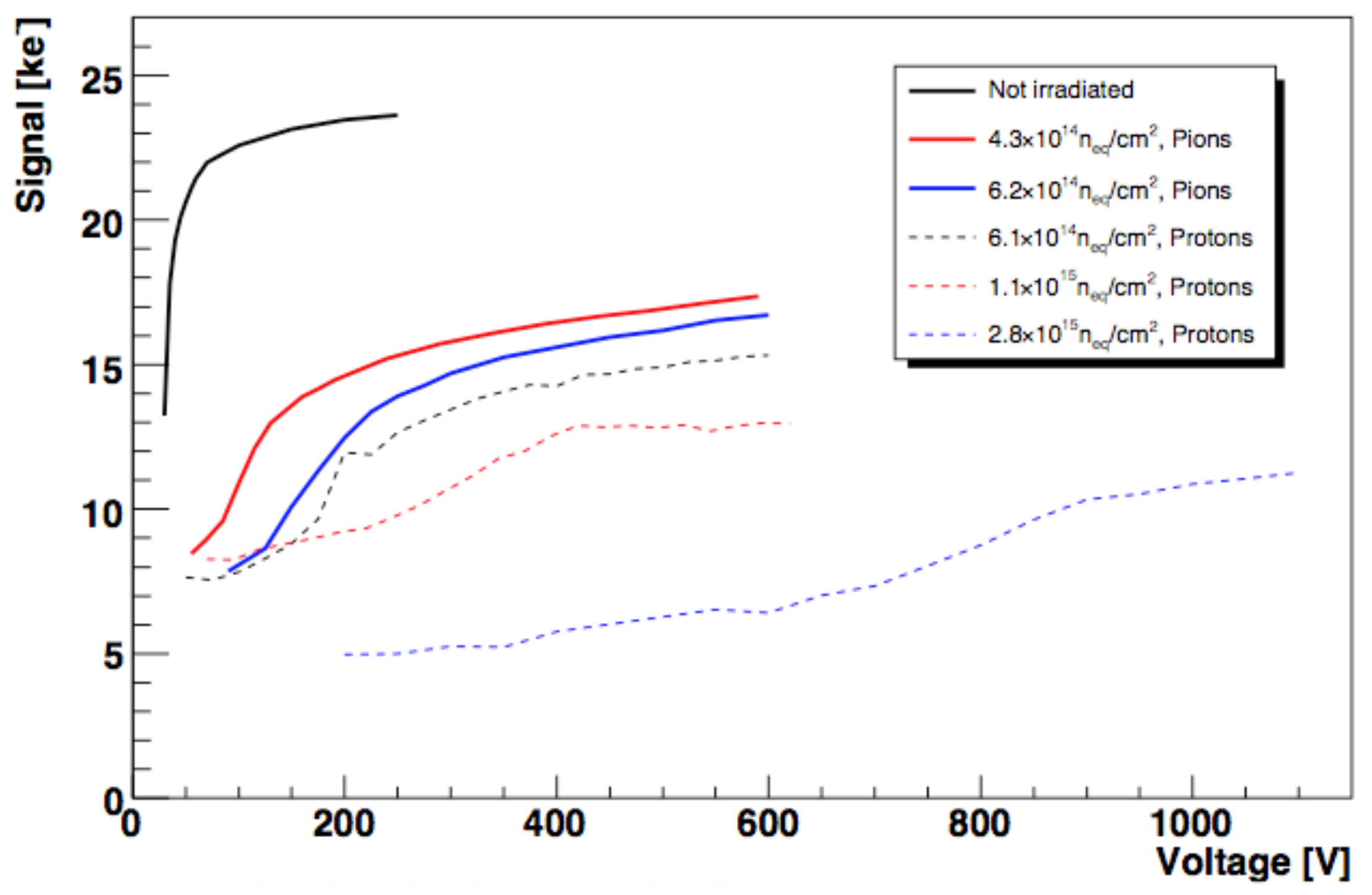}
\caption{\small{Collected charge as a function of the applied bias voltage for various radiation fluences, for the barrel. After a dose of 1.1$\times$10$^{15}$ n$_{eq}$ cm$^{-2}$ and for a bias voltage of 400 V only 50\% of the original charge is collected \cite{radhardnessBPIX}.}}
\label{rad}
\end{center}
\end{figure}
The results are shown in Fig.\ref{rad}. After a dose of  about 10$^{15}$ n$_{eq}$ cm$^{-2}$ only 50\% of the charge is collected, assuming that the bias voltage is raised from 150 to 400 V. This will lower the single hit efficiency below 97\%. In addition, at higher bias voltages the Lorentz drift induced by the magnetic field will be smaller. The resulting reduction of charge sharing between neighboring pixels will worsen dramatically the spatial resolution, which will be dominated by the binary value of single-pixel clusters. After a dose of 1.2$\times$10$^{15}$ n$_{eq}$ cm$^{-2}$, and at a bias voltage of 600 V, the transverse hit resolution will be about two times worse than for the unirradiated sensor. 
\begin{figure}[htbp]
\begin{center}
\includegraphics[height=4.5cm]{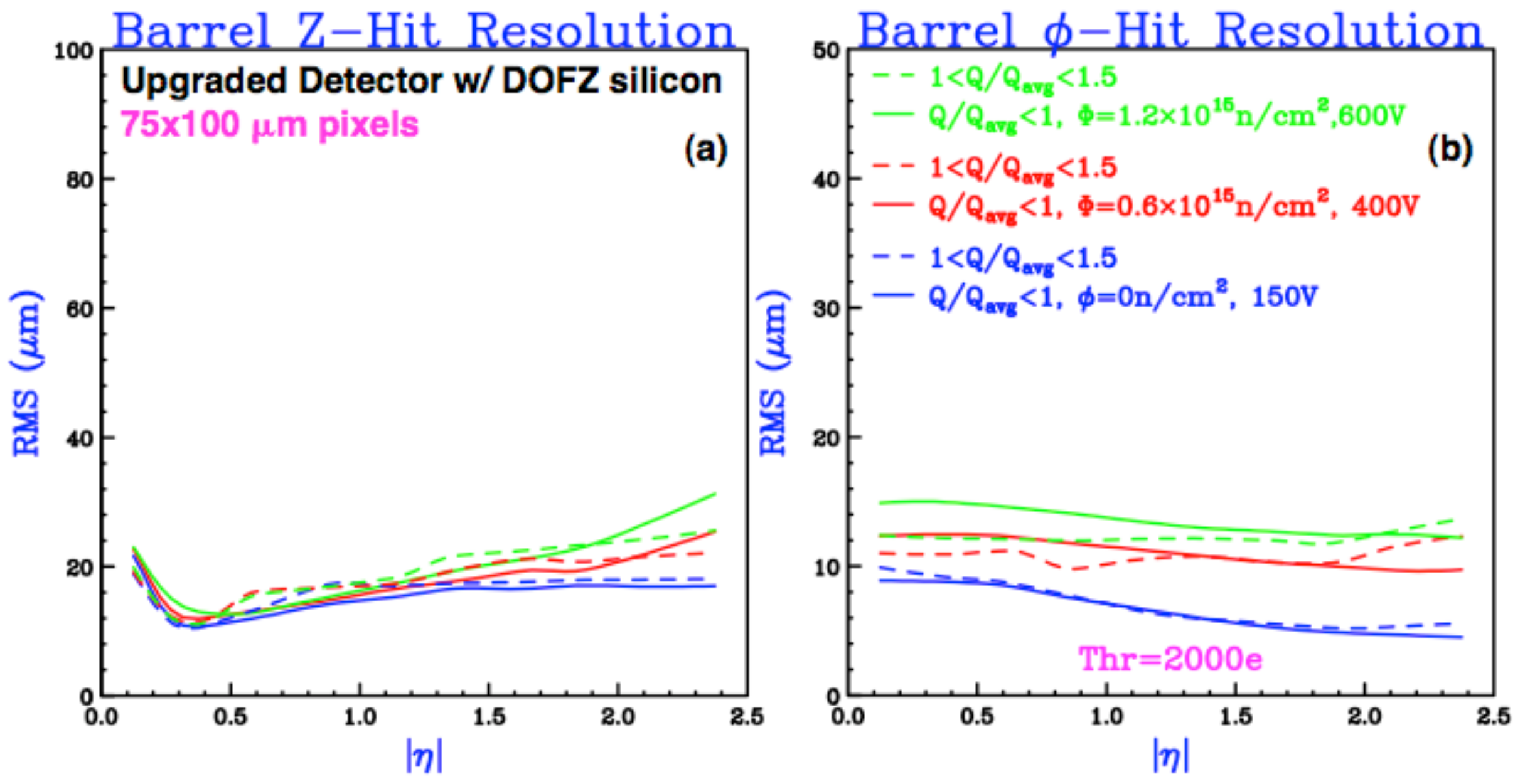}
\caption{\small{Longitudinal (a) and transverse (b) hit position resolution (RMS) for the barrel detector as a function of the track pseudorapidity, for a pixel pitch of 75$\times$100 $\mu$m$^{2}$, a sensor thickness of 200 $\mu$m, and a 2000 electrons readout threshold. Various irradiation scenarios are considered. A detailed simulation of charge deposition and transport in the silicon sensor was used \cite{pixelav}. }}
\label{hitres}
\end{center}
\end{figure}
The innermost part of the detector will need to be replaced before the end of Phase 1, in order to assure good performance throughout the data taking period. The baseline upgrade plan foresees the replacement of the damaged parts with spare components. On the other hand, the opportunity for an enhancement in terms of detection efficiency, spatial resolution, and radiation hardness can be exploited. 

Both the silicon sensor and the frontend readout electronics can be improved. 
The use of mCz silicon instead of the present FZ would increase the sensor resistance to particle fluence, assuring at the same time a similar signal charge collection at low bias voltage. Other options are available and currently under evaluation. \\
For the readout, the possibility to move from the current 250 nm to 130 nm CMOS is considered. This would allow the implementation of a smaller pixel cell, and possibly lower thresholds.  The performance of the innermost pixel layer with 75$\times$100 $\mu$m$^{2}$ pixel cells and a readout threshold of 2000 electrons has been evaluated. The advantage will be a better hit position resolution, lower than 10 $\mu$m in the transverse plane in the full pseudorapidity acceptance, as shown in Fig.\ref{hitres}. The deterioration from irradiation will also be reduced.
\begin{figure}[htbp]
\begin{center}
\includegraphics[height=5.2cm]{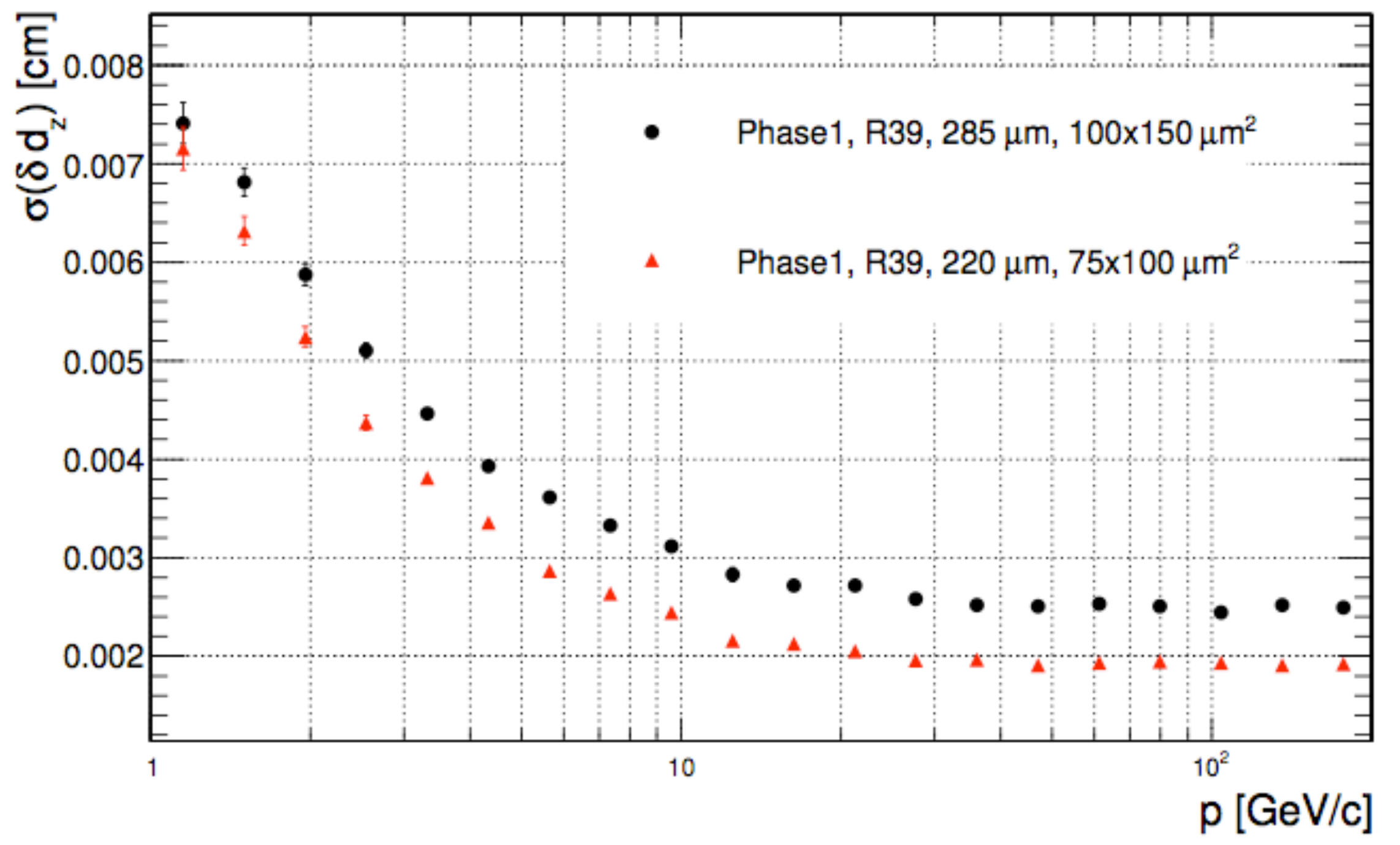}
\caption{\small{Longitudinal impact parameter resolution as function of track momentum, for the baseline Phase 1 barrel detector (in black), and for a hypothetical setup with a first layer implementing smaller pixel cells and lower readout thresholds (in red). The improvement is about 40\% at high momentum. The result was obtained with full simulation.}}
\label{IP_smallpitch}
\end{center}
\end{figure}

The spatial resolution in proximity of the primary interaction region is crucial for the measurement of the track impact parameter. A sizable improvement is thus expected, especially at high momentum, where the multiple scattering effect is negligible. Fig.\ref{IP_smallpitch} shows the impact parameter resolution of the baseline upgraded pixel barrel and a scenario with a first layer with 75 $\times$ 100 $\mu$m$^{2}$ pixel cells, thinner sensors (220 $\mu$m) and lower readout thresholds (2000 electrons). The improvement is about 40\%. The implementation of smaller pixels will also enhance the performance of the $b$ and $\tau$ jet identification at high transverse momenta (above 200 GeV). 

\section{Conclusions}
\label{conclusions}
This article presents the plan for the Phase 1 upgrade of the CMS pixel detector, expected for 2016. The installation of a new detector with enhanced features will be needed, due to the necessity to provide sufficiently good performance in the high luminosity environment. A modified geometrical layout, with a reduced amount of passive material within the active tracking region, a new cooling and a new powering system will be implemented. The readout will be adapted to cope with the higher data rates and prevent data losses. These modifications will provide a significant improvement of the tracking, vertexing and $b$-tagging performance. 

A review of the proposals currently under evaluation for a further development of the innermost part of the pixel detector for late Phase 1, is also given.













\end{document}